# Recent Developments with the licas System 2

Kieran Greer, Distributed Computing Systems, Belfast, UK.
http://distributedcomputingsystems.co.uk
Version 1.1

*Abstract* - The licas (lightweight Internet-based communication for autonomic services) system is a Java-based open source framework for building service-based networks, similar to what you would use a Cloud or SOA platform for. The framework comes with a server for running the services on, mechanisms for adding services to the server, mechanisms for linking services with each other, and mechanisms for allowing the services to communicate with each other. The general architecture of the system is now fairly well set, where this paper describes recent developments that have focused on making the framework more robust and additional features for easier programming.

## 1   Introduction

The licas system [9] is a Java-based open source framework for building service-based networks, similar to what you would use a Cloud or SOA (Service-Oriented Architecture) [8][11][16] platform for. The framework comes with a server for running the services on, mechanisms for adding services to the server, mechanisms for linking services with each other, and mechanisms for allowing the services to communicate with each other. The distributed communication capabilities are now quite flexible. The default communication inside of the system itself is an XML-based RPC (Remote Procedure Call) message passing mechanism. With this, a Java object is parsed into an XML-based description, sent to the remote server and then parsed back again. The call also targets a specific method. There are default parsers provided for most basic types and so you would only need to add a new parser for your own complex Java objects. For building applications however, it might be easier to use a Web Browser and so it is now possible to invoke either HTTP GET or REST-style messages on the server as well. This can even be directly on a local file store, to retrieve HTML or binary files for example, without invoking any service. The capability of using the server simply as a web server is still limited, but it can return some content, similar to what the more basic mobile platforms can provide. Dynamic Web Services invocation through WSDL and SOAP [2][3] is also possible and every base service is loaded with an Autonomic Manager [7][12] that you can add functionality to. The framework is still lightweight and the architecture and adaptive capabilities through dynamic linking, add something new that is not available in other similar systems. The download packages include some AI algorithms and text processing, and an All-in-One GUI can be used either to test systems, or as a practical platform on which to run your own real system. The licas server and related functionality can therefore be used in one of two different ways. Because of resource constraints, these have been packaged together into the one GUI application. The system is also peer-to-peer, with the client GUI also acting as a server that can be invoked.

The rest of the paper is organised as follows: Section 2 gives a brief overview of the architecture again. Section 3 describes the new communication interfaces. Section 4 describes the new support for writing and using services. Section 5 describes the new admin support from the GUI. Section 6 describes the autonomic management framework. Section 7 describes search and organisation possibilities that are provided as default. Section 8 introduces the problem-solving capabilities and section 9 introduces the mobile application possibilities. Section 10 gives some final statements or points.

## 2   Architecture

Licas is a distributed system with a lightweight HTTP server running at its base. The server receives all incoming requests, processes them and sends the message to the indicated service. A service is defined by a path that includes the ip address of the server and then the service name. Security is provided by the fact that passwords are required to invoke a method on the service. If the service message does not contain the appropriate password, it will not be passed on to the related service. The communication mechanism checks this, by asking the service if the password is legal, before invoking the method on it, with the client's data. Any service can also use the communication mechanism to send a request to any other service running on the system. If this is local, then it might be through a direct reference, or if it is remote, then it can be through the XML-RPC mechanism, for example. The licas system is peer-to-peer, but uses this hybrid approach. The process is hybrid, because the request goes from service to remote server and then onto the remote service, but it means that each service or node can both ask for and receive requests.

The linking mechanisms have been written about in detail previously [4][5] and have not changed significantly since the original software versions. There are three different types of link. Firstly, they allow for permanent links between nodes that can define a permanent network structure. They also allow for dynamic links that can change and would reflect more the current system use. There are then also service associations that might not be related to any particular structure, but are just to link any two services through a direct call. There is also the possibility of simply nesting services and allowing direct access through that. It is also relatively easy to extend the base classes to add new functionality. Some work has been done adding basic control or logic, through methods and interfaces, where the user would need to only implement, possibly one new methods from 4 or 5, to create a new service. This includes the autonomic processing and control loop. The system has been written to be J2ME compatible, where Oracle's version of Java [13] is constantly changing in that respect. A restricted version of Java 8 would have problems with the Web Services, for example, but the mobile or embedded versions still exist.

## 3   Communication

A distributed system is composed of a number of individual components that interact with each other, sometimes over a network. The key to the distributed nature is the fact that processing is not centralised in one place (no central control), but is distributed to the different components in the system. Because of this, a global picture can only be obtained if these components can then communicate with each other. A communication mechanism is therefore required, where by default, licas uses an XML-RPC (Remote Procedure Call) style. That is: code any communication into an XML-based script and send this to the requesting component. XML provides a standard format and also a tagging system, so that specific parts of the messages can be labelled and understood. This is also what Web Services use, but in a slightly different format. The format is very safe because everything is clearly defined inside each tag, but it is not the most convenient to type in manually. The server also allows the invocation of a service from a web request. To process that, it accepts REST-style messages. These are full descriptions of the method to call, but coded into a single string, as part of an HTTP address, for example. If the server is not able to parse the message that it receives into the standard 'MethodInfo' object using the default parsers, it will read the request as a REST-style message and try to construct the method call from that. This means that AJAX communications from a web page, for example, are also possible. Figure 1 is a schematic of the different types of communication protocol that are available.

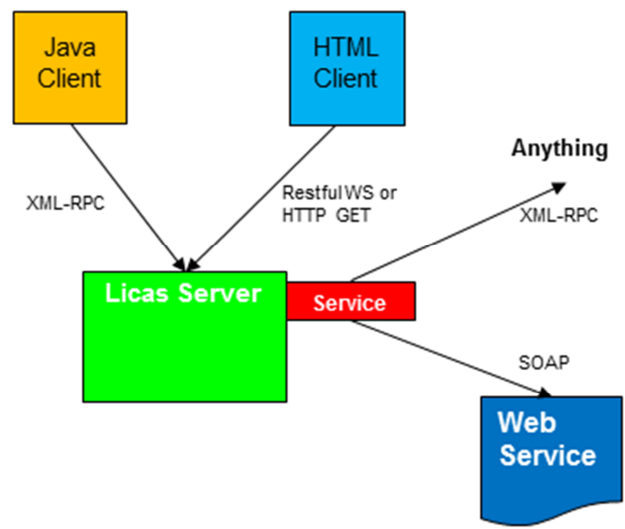

Figure 1. The different communication protocols.

To make some types of call even easier, you can send a message exactly as an HTTP request, without additional REST-style parameters and retrieve either HTML pages, or binary files or documents, directly from the server. This is usually coded as a GET message and some services can also accept a GET request, allowing them to integrate easily into the browser-based client-server architecture. There is also security on the server to make sure that only certain directories can be accessed. So a Java client can make XML-RPC requests. A browser-based one can use HTTP GET or RESTful messages and the server can reply to any of these with the same protocol. The base package also provides for dynamic SOAP messages and while the licas server is not setup to accept these, any service can use the package to make dynamic Web Service requests using WSDL and SOAP. One of the example services can make these requests for you, see section 3.2 and 4.1.

## 3.1 Agent-Based Communications

Agent-based systems [17] can use a more sophisticated communication format that includes pre-defined interaction protocols, so that 'conversations' can take place. Agent-based systems are closely related to autonomous ones or autonomic computing, through the autonomous decisions that they make. An 'Auto' class can provide a base set of functions on which to build agent-based or autonomous systems. It allows for a very free style of conversation, but it also provides the default control loops for executing autonomous behaviours or processes. This means that you only need to add or change one or two methods, rather than write the whole control loop yourself. The possibility of wrapping any Java object, not just ones derived from the licas classes, provides the possibility of using legacy code as agents as well.

## 3.2 Web Services

The licas package includes a Web Services WSDL parser, with the capabilities to dynamically create a SOAP message and to invoke a Web Service from that construction. The same process of parsing and conversion into an XML-based object is applied, only the object that gets parsed is different. It might typically be used as part of a service, where a user can select and call the Web Service method in real time. The Web Service invocation itself is therefore dynamic, where the specific method call itself does not need to be hard-coded beforehand. This means that there is little difference between XML-RPC and Web Services, apart from the exact protocol.

## 4 Services Support

A default All-In-One GUI is provided and can be downloaded from the main DCS page [10], rather than the open source page [9]. It is not open source, but it is free for non-commercial use and provides a lot of helpful functionality. The GUI allows you to load and run your own services, view and interact with them. It includes a licas server that can host its own set of services, as would be the case with a SOA [11] or Cloud [16] system. The GUI can also link or register with other servers and view those configurations as well. So a group of distributed servers can be registered, linked and viewed, and then each individual server configuration can also be viewed. Figure 2 shows the default GUI running a single Web Service service, with the associated service interface also being used. The GUI includes graphic representations of both the server network and the individually selected server configuration. The services can be described through some metadata only and popup menus can be used to interact with them. This includes interacting with remote servers. The GUI therefore provides a visualisation of any network constructed using the licas classes and allows you to test certain functionality as well. The GUI only displays a view of the network, which is based on the metadata descriptions that it retrieves from the server. These metadata objects contain limited XML-based descriptions of the network components. As the view is independent of the service however, this means that it should be possible to create your own service types and still be able to view them.

The HCI interaction with a service can also be made much easier, by allowing it to have its own GUI interface. In Figure 2 it is particularly useful, because calling a Web Service requires it to be initialised with specific values, even if the call itself can be made dynamic. In the context of an SOA therefore, each service can be loaded onto the server and its own GUI opened, as a sort of separate application. The services themselves might be independent of each other, but would still need to be managed and require admin. For testing complex networks, the GUI can display the different link sets that have been created, or return specific data instances. There is also the possibility of opening a separate graphic interface there, to display other specific views. The All-in-One GUI runs on J2SE and therefore has access to all of the Java features. The server itself is written to run on the J2ME CDC subset. It could therefore be used in other programs that use a restricted version of Java.

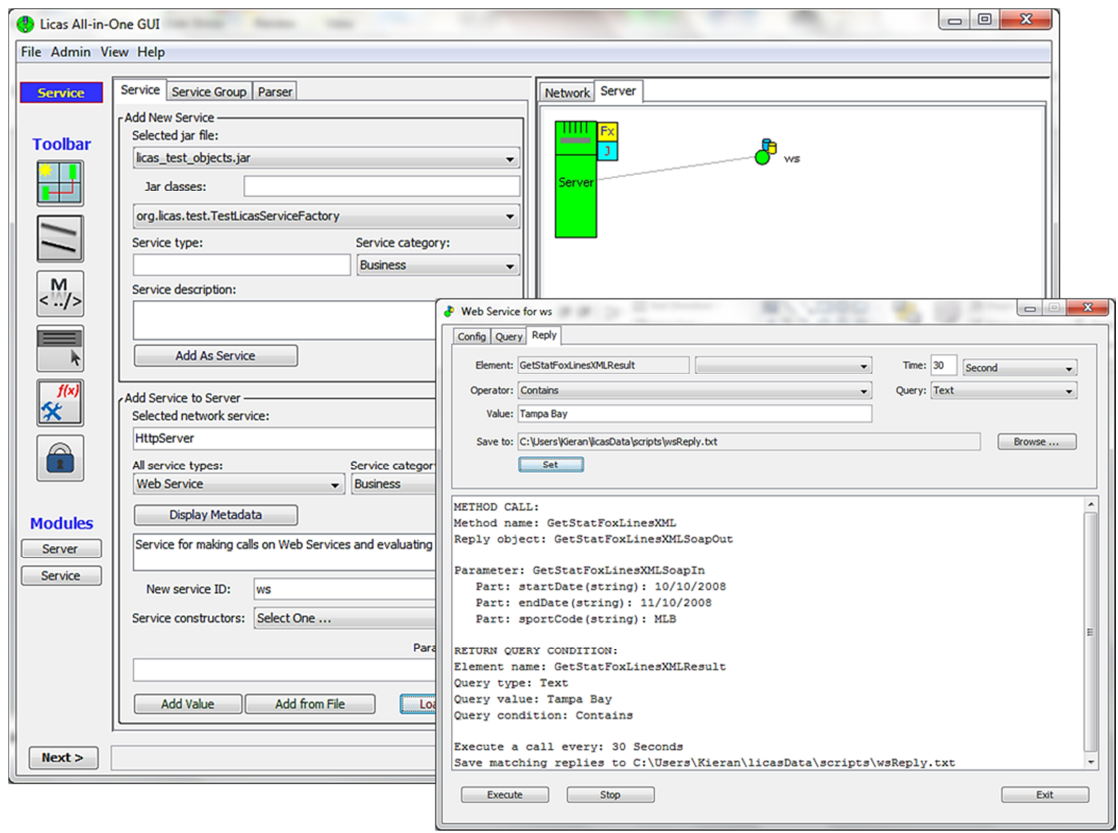

Figure 2. All-In-One GUI, server panel.

## 4.1 Example Services

The download provides three full services as examples. While not perfect, they are still quite useful and might be used as is. There is the Instant Messenger, a p2p File Service and a service for invoking Web Services. The Instant Messenger provides that standard functionality there, including an address book. The File Service is also a standard download from a remote site, although smaller files would be preferred. This is different to a similar GET request from a Web Browser, but you can also browse the files and directories through the GUI interface, for example. The service provides some protection for the directories that you can read from. The Web Services service will allow you to run the request periodically and also set a query condition, so that only replies that match a request get saved. A fourth Information Service is quite interesting. It allows you to reference information resources and has easy methods for retrieving their contents.

## 4.2 Resources

Resources are objects that either store, or link to, information. The idea is that you create a resource to access certain content, where the resource is more lightweight than a service. They are also possibly lightweight versions of the resources that autonomic components might use [12]. In that sense, resources interface with the environment and are used to both receive and send information to or from the autonomic component. The default package provides text or image-based ones, but they could be thought of as linking to any type of external source. The resource also has a standard programmable interface, through which the user can ask for the information. Resource containers can be used to store a number of resources in a single service and again, perform certain pre-defined operations over them. This is may be easier, or more compact, than starting up $x$ services to access $x$ sources, for example. Also, while a service is licas-specific, as a resource can access online content, this give more options for using the rest of the web as part of a program. The Information Service works well with resources, as it has a similar interface for retrieving their contents.

A single resource stores information for a specific type of source. It would typically extend the base 'Resource' class and there are a number of already implemented ones. At the moment, you can store basic types of numbers or strings, and then also XML, HTML, binary, or URL addresses, to remotely access the content. The information can be stored in the resource itself, or retrieved remotely and dynamically. There are also resource containers that store lists of any type of resource and can then return the resource information in a number of pre-defined ways. For example an 'IdContainer' allows you to retrieve the resource info for a resource as specified by its ID. A 'RandomContainer' will select a resource at random and return its contents. A 'QueryContainer' allows you to specify a query as defined by the default 'licas_text' package and will only return the resource content that matches with the query. Containers are created and initialised in the same way as a single resource. The admin script can include each single resource that should be added.

# 5   Administrative Features

A lot of work has also gone into the All-in-One GUI, so that it is more user-friendly and can be used even for real applications. One application would be to support a SOA platform, while the other is for running scientific experiments. To help with settings things up, some new forms that include a number of administrative features have been added to it. For example, a service factory can be used to configure service information or external modules. Specific setups can be saved and automatically re-loaded during start-up, where the details are saved as XML-based scripts. Other scripts can also be used to save details about problem solving tasks, again to allow for the automatic re-execution of the task.

## 5.1   Service Factory

A relatively new feature of the GUI is a 'Service Factory' that can help with most of the service types' setup. This can be opened and used to configure what services can be loaded and also what classes are used. If not changed, the system runs with the default settings. You can also specify or load in external modules, or external Jar files, which can be used to add new services to the lists. After this is saved to the configuration file, the new services are automatically available every time the GUI is run. Figure 3 shows the service factory form. The service factory can also link a windows-based interface to a service, as shown in Figure 2. If the small 'ws' service icon in the graphic area of that figure is right-clicked, a GUI specific to it can be opened if one is registered. The default services have default interfaces, where the figure shows what the Web service one looks like. This just makes interacting with the service much easier. There is also a 'Service+Function' section to allow you to add service types or functionality for the scientific testing.

## 5.2   SOA

There are menu options to allow you to save the configuration of the server that is currently on display. This should normally be the locally running server. The service descriptions are parsed and saved to an XML file set. The file set can then be read in again, to load the service-based descriptions back onto the server. The information that can be saved is still limited, but you can save what services are running and the links between them. You can also save their passwords. You can also do this with the set of registered servers and the links between them. This part is still a work in progress, where additional metadata could be helpful to the re-initialisation process.

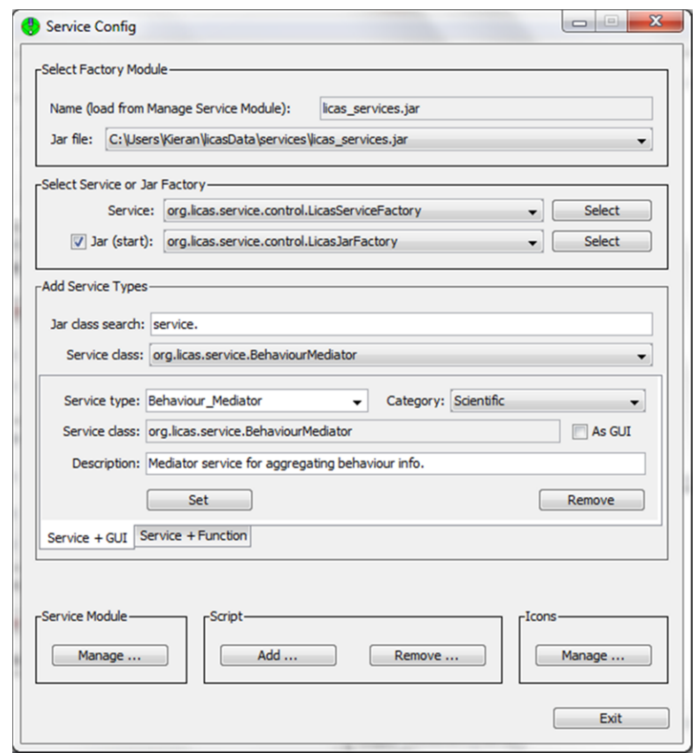

Figure 3. Service Factory form.

## 6  Distributed Autonomic Processing

A distributed system would really benefit from autonomic computing, particularly large-scale systems that need to monitor themselves. Autonomous can mean possibly one of two things. The first meaning would be simply that a component can perform its own operations independently. It has its own internal control system and understanding of what it is required to do. The second meaning is based more on a reactive system with sets of self-features, based loosely on the human nervous system [7]. These features operate automatically and in the background, without prompting and monitor the system, to ensure that it is able to deal with potential faults or threats. There is a Monitor-Analyse-Plan-Execute (MAPE) loop [12] that can typically be implemented, where the system input is fed through these four evaluators that then determine how best to deal with any potential threat. There are still great technical problems with building a system that would be 100% autonomous as this would also require the system to also be very intelligent and understand many different types of input. This can be helped by allowing a system administrator to define rules for the system through admin scripts, or by simply limiting the level of intelligence that the system is required to use. Figure 4 is a schematic of the general framework and architecture of an autonomic service provided by licas.

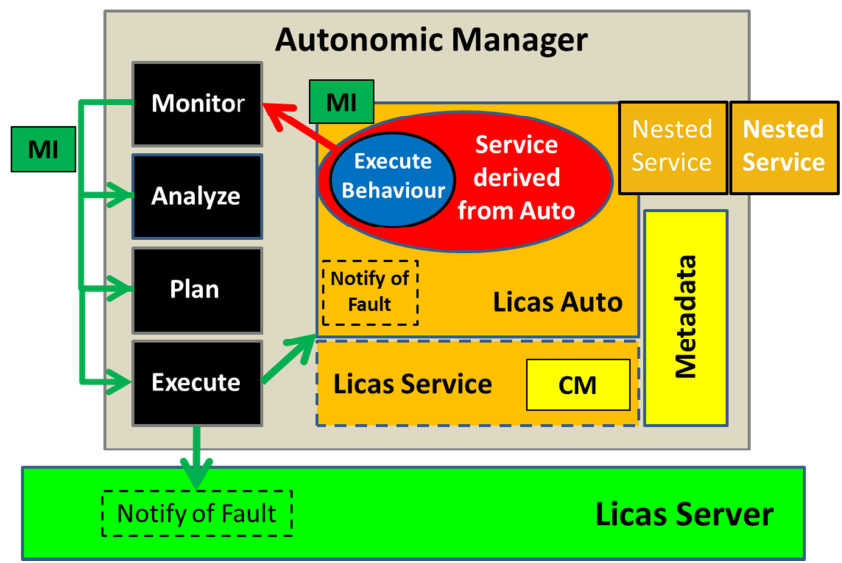

Figure 4. Schematic of the licas service architecture

A set of derived service classes are shown in the central orange-red squares, with the licas Service at the bottom. In the figure, the base class is a 'Service' class. There is default metadata relating to the service and also a Contract Manager, to store and reason over proposed contracts for the service work. Nested or utility services can be added to any other service, but should in theory be used more as an aid, rather than fully independent services themselves. For autonomic components, the derived 'Auto' class is required. You can then extend either of the base classes with a new service implementation. In the figure, the red oval is a service that extends Auto and executes a behaviour. As default, the system adds an Autonomic Manager to every base service that gets registered with the server, or the wrapper that encapsulates it. The Autonomic Manager includes slots for the 4 modules – Monitor, Analyze, Plan and Execute, but they are left empty. It is up to a programmer to add actual implementations that can be configured through a script. There is also a Message Interface (MI). On executing a behaviour in the service, the Auto class will automatically send the result to the Autonomic Manager, using the message interface. If autonomic modules have been implemented, they can then process the input. If there is a fault, both the service itself and the base server can be informed. The resources of section 4.2 might even be considered as lightweight versions of the equivalent autonomic managed resource concept.

# 7   Search and Organisation

For large-scale distributed systems, there is a need to organise them efficiently, to optimise for time and resource usage. The individual components also need to be found and ideally, be able to link or combine with each other, to allow for more complex tasks and interactions. The system has not been tested to this level or scale, but it is possible that these problems could arise. The service management is known as service Deployment, Discovery and Composition [6]. Theoretically, using licas, individual services can be loaded at run time and coded to perform any sort of action. They can be configured through scripts and can communicate and link with any other service. There is also a query language and service-specific metadata, to allow for the search operations. The query language and engine are provided as part of a text package. It is not open source, but is part of a set of utility packages that licas also uses. It does allow for processing of XML-based, or simply text-based data and uses a new type of query description. While the language is not a standard, it is still intuitive enough. If you are only concerned with defining what information you wish to find, you can use it relatively easily. It can define path patterns and constraints on those patterns that determine what information gets returned; or simply, different types of evaluation on lines of text. So the licas system already provides for descriptive metadata and query operations that would allow for relatively sophisticated query processing over distributed networks.

# 8 Problem Solving

The licas system is both distributed and provides a framework for adding autonomic monitoring components. The communication language allows individual nodes to talk to each other, allowing them to perform evaluations based on local information, typically between two services only. The other classical setup is to bring all of the information together into a centralised place. In that case a more global and complete picture can be obtained, but the processing requirements are much more heavyweight. Licas allows for a centralised problem solver to be run. This can request or receive information from the services running on its network, process them, including AI tree searches for solutions and then send the results of this back to the individual nodes. So both the lightweight localised processing and the more heavyweight centralised processing can be accommodated for. The individual servers can also link up, making scalability to larger systems possible. One intention is for the system to still be useful as a research tool and so the problem solving aspect is particularly important there.

The problem solver can be configured with its own script, where the different variable values can be entered through the GUI. The GUI also allows the default service types to be updated or changed. The default system provides a 'Behaviour' service and an 'Information' service, and a mediator to manage the interactions. The behaviour service can be extended by programming only one or two methods and will then execute the behaviour of those services periodically and return the results for processing. The default problem-solving mechanism is genetic algorithms, where text-based information is sent to the central problem solver, processed through evolving genetic solutions and the result returned to the distributed services. There is now also the distributed linking option, where localised evaluations between services determine what services link with each other. The download package includes user manuals that describe each of the system applications and the problem solver can also be run in a Command Prompt box, if a network structure is not required.

## 9  Mobile Applications

The Oracle Java Mobile environment has been changing. Currently, there is a Java 8 version [13] that includes restricted versions of the JVM. These are intended for restricted versions of the Java language, but licas includes the Web Service SOAP classes that are only available under the full Java 8 version. The mobile platforms still exist however and include Java 8 Me Embedded. Licas has been built using Java ME CDC with some additional packages. Mobile devices tend to require a Midlet to run, which means Java CLDC. The embedded devices might allow CDC however. For a self-contained installer however, an early version of a J2SE JRE [15] could be included, such as Java 1.4, and would keep the download size to a minimum. So it is as much a problem of compatibility with the device, than with supplying all of the required Java classes. However, JavaFX [14] has recently announced that it may become mobile compatible again in the near future and other options are also attractive. A lightweight server has been written and is available from the web site [10]. It could be used along with a HTML-based client, for example, to provide an access point to the Internet, removing some of the application processing from the mobile phone. For this, Cordova [1] looks like a possibility. So there are other options, including simply HTML in a Browser that can provide a satisfactory client from which to use the licas services remotely. Tablets should already be compatible as well. Porting to Linux also looks possible, where external installations and variables simply need to be set. So a complete mobile client-server package is still problematic but certainly looks possible.

## 10  Discussion

Recent work has focussed on making the system more robust again, so that it might be used in real-world scenarios and not just as a test platform. However, the research-side has also been kept, as is it a flexible system that is ideal for certain distributed or dynamic network tests. There have been real improvements on the framework side, with regard to how a user might add autonomic processes. The user now only needs to implement one or two additional methods, to allow for the autonomic monitoring or processing of information, for example. Security has also been improved with passwords, contracts and a wrapper to protect direct service references. The web service parsers and now quite robust and the web service invocation has been integrated into the system, as part of the whole communication mechanism. Resources look like a useful addition and open the system up to the rest of the Web. Finally, system configuration has been tackled, with a service factory and configuration scripts, allowing for SOA configurations to be saved. These can then be re-used with minimal effort to the user.

A lightweight framework, such as that provided by licas, offers a nice solution to running your own server. Your mobile device is usually on and connected to a network, but would probably be a tablet for the Java server. You would not typically host static web pages on your phone, but might process location-based or information-based data, in a much more dynamic way. The system therefore also needs to be helpful, when autonomic processing is useful. Scale has not been tested, but the licas system might be used to build more localised networks between small numbers of people that do not flow through the major vendors' systems. This might even help with network congestion problems. The problem at the moment is the compatibility issue with the main mobile operating systems. Development for mobile devices would be easier if the CDC platform became easier to use, but things are developing in this respect and there are currently some good options. This is not the main goal of the project however, which is to provide the server only, but would be an interesting addition.

## Acknowledgements

This is a second update, following the paper 'Recent Developments with the licas System', published on Scribd in 2010 and a second version of this paper, first published in 2013.